\documentclass[11pt,a4paper]{article}
\usepackage{fullpage}
\usepackage{amsfonts}
\usepackage{amssymb}

\title{Calculating few-body resonances using an oscillator trap}

\author{
D.V.~Fedorov, A.S.~Jensen, M.~Th{\o}gersen\\
{\it Aarhus University, 8000 Aarhus C, Denmark}\\
E.~Garrido, R. de Diego\\
{\it Instituto de Estructura de la Materia}\\
{\it CSIC, Serrano 123 E-28006 Madrid, Spain}
}
\date{}

\begin{document}

\maketitle
\begin{abstract}
We investigate the possibility of calculating the parameters of few-body
resonances using the oscillator trap boundary conditions. We place the
few-body system in an oscillator trap and calculate the energy spectrum
and the strength function of a suitably chosen transition. Broader
resonances are identified as Lorentzian peaks in the strength
function. Narrower resonances are identified through the pattern of
avoided crossings in the spectrum of the system as function of the
trap size. As an example we calculate $0^+_2$ and $0^+_3$ resonances in
$^{12}$C within the 3$\alpha$ model.
	\end{abstract}

\section{Introduction}

Few-body resonances are often calculated using the complex scaling method
where the resonances are identified as generalized complex eigenvalues
of the Hamiltonian with the corresponding generalized eigenfunctions
(see e.g. \cite{elander,fed-complex} and references therein). The method
has the advantage of having a simple boundary condition: the few-body
wave-function vanishes at large distances.  However, it also has certain
disadvantages. Complex arithmetics and algorithms are generally slower and
complex matrices need more computer memory. Calculating extremely narrow
resonances is difficult as it demands calculations of the eigenvalues
with exceedingly high accuracy.  Interpretation of the generalized
eigenfunctions is also not trivial \cite{elander}, especially for heavy
complex scaling needed for broader resonances.

In this contribution we investigate the possibility of calculating
the parameters of few-body resonances using the same simple boundary
condition as in complex scaling method but working with real energies
and real wave-functions.

We place the few-body system in an artificial oscillator trap of length
$b$ which is significantly larger than the characteristic length of the
few-body system. We then calculate the (discrete) spectrum of the system
in the trap and estimate the strength function of a certain transition
from a suitable chosen initial state to the positive-energy states of
the system in the trap.

The broader resonances with width $\Gamma\gtrsim\hbar^2/(2mb^2)$, where
$m$ is the characteristic mass of the few-body system, can be identified
as Lorentzian peaks in the strength function.

The narrow resonances with $\Gamma<\hbar^2/(2mb^2)$ need an investigation
of the spectrum of the system in the trap as function of the trap
length. When an energy level in the trap, following its general behavior
as $b^{-2}$, approaches the system's resonance level to within its width,
the two levels interfere and avoid crossing.  The pattern of avoided
crossings in the spectrum of the system in the trap reveals the position
of the narrow resonance. The width of the resonance can be estimated
from the size of the region of avoided crossing, or, more precisely,
from the variation of the energy levels with respect to the trap length.

The approach is similar to the box method (also called the stabilization
method)~\cite{spherical-box,zhang}.  However, the difference is that we
use an oscillator trap instead of a box and that we resort to strength
function method for broader resonances where the avoided crossings
method is less reliable. The box boundary condition is more complicated
as the wave-function has to vanish identically at the box boundary which
for few-body systems is a multi-dimensional surface.  The oscillator
trap can be potentially used in stochastic variational calculations with
correlated Gaussians~\cite{martin}.

As an example we apply the approach to the 3$\alpha$ system in the
$J^\pi$=$0^+$ channel where there exist a narrow, $0_2^+$, and a broader,
$0_3^+$, resonance. We show that the approach allows to reliably
calculate the two resonances in this system.

\section{The few-body system and the trap}

We consider the 3$\alpha$ system with the total angular momentum and
parity $J^\pi$=0$^+$.  The Ali-Bodmer type $\alpha$-$\alpha$ potential
is taken from \cite{fed96},
	\begin{eqnarray}
V_{\alpha\alpha}(r)&=& \left(125\hat{P}_{l=0} +20\hat{P}_{l=2}\right)
e^{-\left(r\over 1.53\right)^2} -30.18\,e^{-\left(r\over 2.85\right)^2}
\nonumber \\ &+&{4\cdot1.44\over r} \;\mathrm{erf}\left({r\over
2.32}\right),
	\end{eqnarray}
where all energies are in MeV, all lengths in fm, $\hat{P}_{l}$ is the
projection operator onto a state with relative orbital momentum~$l$,
and $r$ is the distance between $\alpha$-particles.
In addition a three-body force
	\begin{equation}
V_3(\rho)=-76\,\mathrm{MeV}\exp(-\rho^2/(4\mathrm{fm})^2)\;,
	\end{equation}
is employed to simulate the contribution of ``compound nucleus'' degrees
of freedom at shorter distances where all three $\alpha$-particles
overlap. The three-body force is defined in terms of the hyper-radius
	\begin{equation}\label{eq-rho}
\rho^2={m_\alpha\over m}\sum_{i=1}^{3}r_i^2\;,
	\end{equation}
where $r_i$ are the c.m. coordinates of the $\alpha$-particles,
$m$=939~MeV is the chosen mass scale and $m_\alpha$=3.97$m$.

The system is placed in an oscillator trap
	\begin{equation}\label{eq-vtrap}
V_\mathrm{trap}={\hbar^2\over 2m}{\rho^2\over b^4}\;,
	\end{equation}
where the trap length $b$ is varied around 30-40~fm.

The three-body problem in the trap is solved using the adiabatic
hyper-spherical method (see e.g. \cite{fed-complex} and references
therein). First for every fixed hyper-radius $\rho$ the eigenvalue
problem for all remaining variables (denoted collectively as
hyper-angles $\Omega$) is solved and the spectrum of hyper-angular
eigenvalues $\epsilon_i(\rho)$ together with the angular eigenfunctions
$\Phi_i(\rho,\Omega)$ are obtained. The functions $\Phi_i(\rho,\Omega)$
are then used as a full basis in the $\Omega$ space and the total
wave-function $\psi$ is represented as a series
	\begin{equation}
\psi(\rho,\Omega)=\sum_{i=1}^{\infty}f_i(\rho)\Phi_i(\rho,\Omega)\;,
	\end{equation}
where the expansion coefficients $f_i(\rho)$ are obtained by solving the
hyper-radial equations where the eigenvalues $\epsilon_i(\rho)$ serve
as effective potentials.

\section{Strength function}

A resonance can be identified as a peak in a reaction cross-section with
approximately Lorentzian shape.  The amplitude of a quantum transition,
caused by an operator $F$, from some initial state $\psi_a$ into one of
the discrete state $\psi_n$ of the system in the trap, is given in the
Born approximation as
	\begin{equation}
M_{n\leftarrow a}=\langle \psi_n | F | \psi_a \rangle \;.
	\end{equation}
Since a resonance per definition must be seen in any reaction channel,
the particular choice of the excitation operator $F$ and the initial
state $\psi_a$ should be irrelevant as soon as the matrix element does
not vanish identically. We thus choose the initial state in
the form of the large $\rho$ asymptotics of a bound three-body
state~\cite{nielsen,pushkin},
	\begin{equation}
f_i(\rho)=\rho^{-5/2}\exp(-\rho/b_3),
	\end{equation}
in every hyper-radial channel $i$.  The constant $b_3$=4\,fm is chosen
close to the size of the bound state of 3$\alpha$ system. The excitation
operator is taken as
	\begin{equation}
F=\rho^2\;.
	\end{equation}

\begin{figure}[tbh]
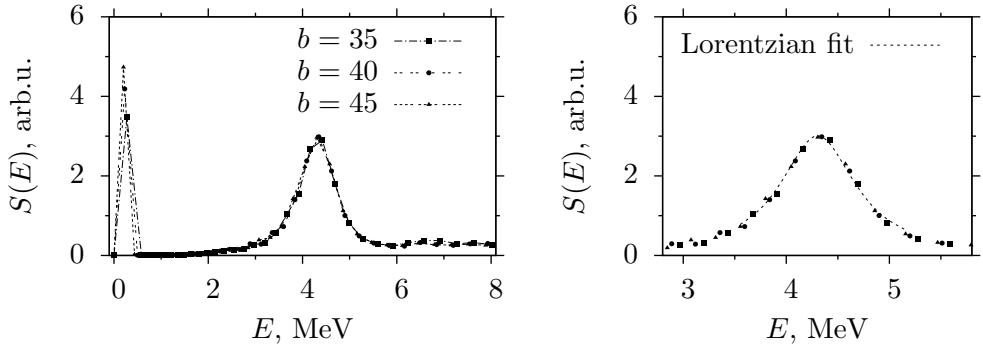

\begin{tabular}{cc}
\input fig-dbde.tex & \input fig-bw.tex
\end{tabular}
\caption{Left: the strength function $S(E)$ as function
of the 3$\alpha$ energy $E$ for different oscillator lengths
$b$. Right: the peak at 4.3\,MeV is fitted with a Lorentzian,
$\mathrm{Const}/[(E-E_{r})^2+\Gamma^2/4]$, where $E_{r}$=4.3~MeV and
$\Gamma$=0.9~MeV.}
	\label{fig-dbde}
\end{figure}

The cross-section of a reaction into the final states with energies
$E\pm{\Delta E\over2}$ is determined by the strength function, defined as
	\begin{equation}
S(E)={1\over\Delta E}\sum_{E_n\in E\pm{\Delta E\over2}} |M_{n\leftarrow
a}|^2  \;.
	\end{equation}
The energy bin size $\Delta E$ has to be chosen on the one hand small
enough as not to smear out the essential features of the cross-section,
and on the other hand large enough to include many states. In our
calculations the energy bins include four states each.

The calculated strength function is shown on Figure~\ref{fig-dbde}. It
reveals a broader peak at 4.3~MeV and a narrow unresolved peak at
$\sim$0.4~MeV. In the region of the broader peak the strength function
is well converged with respect to the trap length, and the bin size is
quite appropriate for the description of the width of the peak as there
are many points within the peak region.

The shape of the peak is well described by a Lorentzian
	\begin{equation}
S(E)\stackrel{E\approx E_r}{\propto}
{1\over (E-E_r)^2+{\Gamma^2\over4}}
	\end{equation}
with $E_r$=4.3~MeV and $\Gamma$=0.9~MeV. These numbers are consistent
with~ \cite{fed-complex}.

The narrow peak at in the strength function $\sim$0.4MeV is represented by
only one point. The position of the point reveals the resonance energy
but not the width. To resolve the width at least several points are
needed within the peak region. For resonances width exceedingly narrow
width $\Gamma$ this would demand unreasonably large trap lengths of the
order $b\sim\sqrt{\hbar^2/(2m\Gamma)}$.

However instead of the strength function the avoided crossings method
can be used to calculate narrow resonances using reasonably sized traps.

\section{Avoided crossings}

For large trap lengths the energy levels in the trap scale with the trap
size as $\hbar\omega\propto b^{-2}$. Varying the trap size a level
in the trap can be moved close to the resonance level of the system.
If the resonance were behind a completely impenetrable barrier (thus
having a vanishing width) there would be no interference through the
barrier between the resonance and the state in the external trap. The
resonance would then be insensitive to the trap size. The spectrum of
the system in the trap, as function of the trap size, would thus show
the trap levels scaling as $b^{-2}$ and crossing the resonance energy
represented by a horizontal line.

\begin{figure}[t]
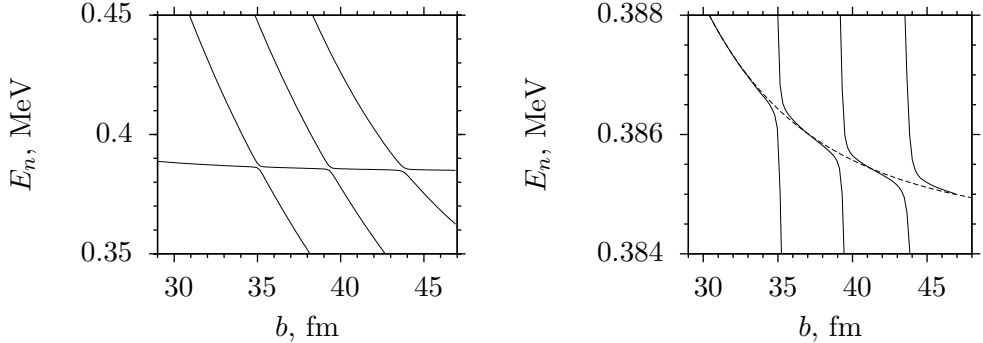

\begin{tabular}{cc}
\input fig-epos.tex & \input fig-e2.tex
\end{tabular}
\caption{Left: the spectrum of the 3$\alpha$ system in an oscillator trap
as function of the trap size $b$ in the region of the narrow peak
on Figure~\ref{fig-dbde}: the sequence of avoided crossings indicates a
resonance at $\sim$~0.38~MeV; Right: zoom-in into the region of avoided
crossings: the resonance energy is fitted with $E(b)=E_r+Kb^{-4}$
with $E_r=0.38435$~MeV and $K$=7.471$^4$~MeV$\;$fm$^4$.}
	\label{fig-epos}
\end{figure}

If the barrier has small but finite penetrability the trap level
approaching the resonance to within its width becomes perturbed by
the resonance resulting in the ``repulsion'' of the two interfering
levels. This shows up as a sequence of avoided crossings in the spectrum
of the system in the trap as function of the trap size.

Indeed the spectrum of the 3$\alpha$ system in the trap reveals such a
sequence of avoided crossings in the vicinity of the narrow resonance,
see Figure~\ref{fig-epos} (left).

The resonance state gets a contribution from the oscillator potential
(\ref{eq-vtrap}) which at large $b$ is proportional to $b^{-4}$. This
contribution can be determined by a fit $E_r+Kb^{-4}$ through the
resonance energies as shown on Figure~\ref{fig-epos} (right). The
fit also provides the asymptotic estimate of the resonance energy
$E_r=0.38435$\,MeV.

\begin{figure}[tbh]
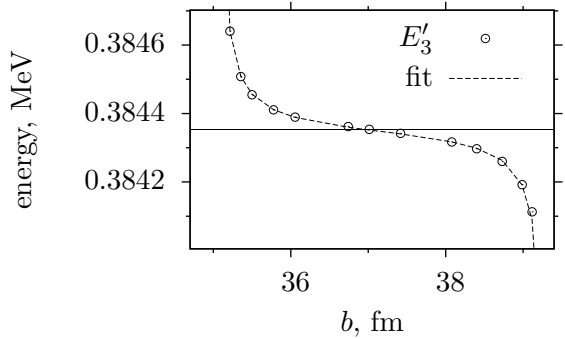

\centerline{\input fig-res.tex}
\caption{The reduced energy $E_3^\prime\equiv E_3-Kb^{-4}$ of the third
level of the 3$\alpha$ system in an oscillator trap as function of
the trap length $b$ in the region of the resonance $E_r=0.38435$\,MeV
(indicated as a horizontal line) where the parameters $K$ and  $E_r$
are from the fit of the resonance energy on Figure~\ref{fig-epos}. The
energy $E_3^\prime$ is fitted with the curve ${b-b_3\over\Delta
b_3}=\arctan{\Gamma/2\over E-E_r}$ where $b_3$, $\Delta b_3$, and $\Gamma$
are fitting parameters. The fit gives $\Gamma=77$\,eV.}
	\label{fig-res}
\end{figure}

The figure~\ref{fig-res} shows the reduced energy $E_3^\prime\equiv
E_3-Kb^{-4}$ where the oscillator contribution is subtracted. It is
possible to estimate the width $\Gamma$ of the resonance from the plot
assuming that the energy region where the avoided crossing takes place
is determined by the width of the resonance,
	\begin{equation}
{\Gamma\over2}=\Delta b\left.{\partial E_n^\prime\over\partial
b}\right|_{E_n=E_r}\;,
	\end{equation}
where $\Delta b$ is the distance between the neighboring avoided
crossings.

However, instead of numerical differentiation it is better to estimate
the width by fitting the calculated energies with the curve
	\begin{equation}
{b-b_n\over\Delta b_n}=\arctan{\Gamma/2\over E_n^\prime-E_r}\;,
	\end{equation}
where $b_n$, $\Delta b_n$, and $\Gamma$ are fitting parameters.
Figure~\ref{fig-res} (right) shows such a fit for $E_3$. The fit gives
$\Gamma=77$\,eV. This value is consistent with the estimates of 10-30\,eV
in \cite{fed96,fedotov} taking into account that our three-body potential
provides a slightly higher $E_r$.

\section{Conclusion}

Using the two lowest $J^\pi$=0$^+$ resonances in the 3$\alpha$ system
as an example, we have investigated the possibility of calculating the
energies and widths of few-body resonances by placing the few-body system
in an artificial oscillator trap.  The oscillator trap has particularly
simple boundary condition and can be potentially used in stochastic
variational calculations with correlated Gaussians.

We have shown that broader resonances with the width
$\Gamma\gtrsim\hbar^2/(2mb^2)$, where $b$ is the trap size, can be
identified as Lorentzian peaks in the strength function of a suitably
chosen ``gedanken'' transition.  Narrower resonances can be identified
through the pattern of avoided crossings in the spectrum of the system
in the trap as function of the trap size.

\end{document}